\documentclass[aps,prb,superscriptaddress,reprint]{revtex4-2}

\usepackage{graphicx}
\usepackage{latexsym}
\usepackage{dcolumn}
\usepackage[utf8]{inputenc}
\usepackage[T1]{fontenc}
\usepackage{tabularx}
\usepackage{hyperref}
\usepackage{color}
\usepackage{placeins}
\usepackage{float}
\usepackage{amsmath}
\usepackage{gensymb}
\usepackage{url}

\begin{document}

\title{Atomistic modelling and structural characterisation of coated gold nanoparticles for biomedical applications}

\author{Matthew D. Dickers}
\email{M.D.Dickers@kent.ac.uk}
\affiliation{School of Physics and Astronomy, Ingram Building, University of Kent, Canterbury, CT2 7NH, United Kingdom}

\author{Alexey V. Verkhovtsev}
\email{verkhovtsev@mbnexplorer.com}
\affiliation{MBN Research Center, Altenh\"oferallee 3, 60438 Frankfurt am Main, Germany}

\author{Nigel J. Mason}
\affiliation{School of Physics and Astronomy, Ingram Building, University of Kent, Canterbury, CT2 7NH, United Kingdom}

\author{Andrey V. Solov'yov}
\affiliation{MBN Research Center, Altenh\"oferallee 3, 60438 Frankfurt am Main, Germany}


\begin{abstract}
This study presents the results of atomistic structural characterisation of 3.7 nm diameter gold nanoparticles (NP) coated with polymer polyethylene glycol (PEG)-based ligands of different lengths (containing $2-14$ monomers) and solvated in water. The system size and composition are selected in connection to several experimental studies of radiosensitisation mechanisms of gold NPs. The coating structure and water distribution near the NP surface are characterised on the atomistic level by means of molecular dynamics simulations. The results of simulations carried out in this study, combined with the results of our recent study [J. Phys. Chem. A \textbf{126} (2022) 2170] and those from the field of polymer physics, are used to calculate key structural parameters of the coatings of radiosensitising gold NPs. On this basis, connections between the coating structure and distribution of water are established for different NP sizes as well as lengths and surface densities of coating molecules.
The quantitative analysis of water distribution in the vicinity of coated metal NPs can be used to evaluate the radiosensitising effectiveness of a particular NP system based on the proximity of water to the NP metal core, which should impact the production of hydroxyl radicals and reactive oxygen species in the vicinity of metal NPs exposed to ionising radiation.
\end{abstract}


\maketitle

\section{Introduction}
\label{sec:Intro}

Metal nanoparticles (NPs) have garnered much attention in recent years due to their wide range of biotechnology and biomedical applications, including medical imaging \cite{Han_2019_Nanoscale.11.799}, drug delivery \cite{Anderson_2019_NRL_magneticNPs}, and radiotherapy of cancers \cite{CancerRadiotherapyReview, Khursheed_2022_metalNPs_biomed_review}. When injected into the body and localised within a tumour region, metal NPs can act as radiosensitising agents through irradiation with x-rays and ion beams \cite{CancerRadiotherapyReview, NP-enhanced-RT_book_2020, Schuemann_2020}.
The interaction of metal (typically gold) NPs with ionising radiation results in the strong emission of secondary electrons \cite{Verkhovtsev_2015_PRL.114.063401, Verkhovtsev_2015_JPCC.119.11000, Chow_2016_chapter}, which interact with molecules of the surrounding medium (that is, to a large extent, water). As a result of such interactions, hydroxyl radicals and other reactive oxygen species are produced, which may deliver damage to tumour cells \cite{HydroxylRadicalProduction, GILLES2014770}.

At the same time, the use of metal NPs for biomedical applications can lead to different cytotoxic effects, including immune responses \cite{NPImmuneResponse}, inflammation \cite{Inflimation}, and damage to cell membranes \cite{CellMembraneDamage}. In order to reduce toxicity, metal NPs proposed for biomedical applications are usually synthesised with organic coatings, which also help to increase the uptake and longevity of NPs within cells \cite{AdverseEffects}. One of the most popular coatings used in experiments with radiosensitising NPs is poly(ethylene glycol) (PEG), often chosen as it increases the stability and reduces the toxicity of NPs injected into biological systems \cite{Karakoti2011-jw}.

Structural and energetic properties of coated (also commonly referred to as functionalised or ligand-protected) metal NPs have been widely studied computationally by means of density functional theory (DFT), classical molecular dynamics (MD), and Monte Carlo methods \cite{doi:10.1021/acs.chemrev.5b00703, C5CP01136A, doi:10.1021/acs.jctc.5b01053, C6CP05562A, Haume2016, Malola2021, doi:10.1021/acs.jpclett.0c00300, LIN2017325, Verkhovtsev_2022, PEO}. In particular, several studies \cite{Verkhovtsev_2022, Haume2016} were devoted to the atomistic MD-based modelling and structural characterisation of coated metal NPs in view of their radiosensitising properties. In the cited studies, gold NPs with a diameter of 1.6~nm were coated with thiol-PEG$_n$-amine (S-(CH$_2$)$_2$-PEG$_n$-NH$_2$) molecules and solvated in water. The structural properties of the coatings and the distribution of water in the vicinity of the NPs were analysed at the atomistic level.

In Ref.~\cite{Haume2018}, the transport of low-energy secondary electrons and the production of hydroxyl radicals in the vicinity of 1.6-nm gold NPs coated with thiol-PEG-amine ligands has been studied theoretically and computationally. It was demonstrated that the production of secondary electrons and OH$^{\bullet}$ radicals in the vicinity of coated metal NPs is highly dependent on water distribution near the NP surface. A greater amount of water close to the NP surface results in a larger yield of hydroxyl radicals formed in that spatial region. As such, determining water distribution in the vicinity of coated metal NPs is essential to quantify the production of secondary electrons and free radicals and, therefore, to evaluate the effectiveness of such NPs as radiosensitising agents for a particular NP size and elemental composition.
The distribution of water surrounding the gold NPs is highly dependent on the structure of the coating, with higher-density coatings preventing water molecules from penetrating close to the NP surface \cite{GILLES2014770, Verkhovtsev_2022}. It is, therefore, essential to understand, on the atomistic level, the structure of the NP coating to predict the distribution of water molecules.

Gold NPs studied in experiments on NP radiosensitisation have a wide range of sizes from $\sim$2 to 100~nm \cite{Sophie}.
For NPs with a size of several ($\sim$1--10) nanometers, the structural characterisation of the metal core, coating, and surrounding molecular environment can be performed using the all-atom MD-based approach \cite{Verkhovtsev_2022, Haume2016}.

However, as the system size increases, this approach becomes computationally expensive and even infeasible for NPs with a size of $\sim$10 nm and above, considered by some experimental studies \cite{Sophie, HydroxylRadicalProduction, GILLES2014770}. A more viable approach is to develop a model for describing and predicting the structural properties of coated metal NPs, such that the coating structure and the distribution of water molecules near the metal NP surface can be analysed for any NP size as well as for any length and surface density of ligand molecules.

This paper aims to take a further step in this direction. In the present paper, the structure of an experimentally relevant system -- a gold NP with a diameter of 3.7~nm functionalised with thiol-PEG$_n$-amine $(n = 2-14)$ molecules and solvated in water is characterised on the atomistic level using the MBN Explorer \cite{MBNExplorer} and MBN Studio \cite{MBNStudio} software packages. Based on the results of structural characterisation carried out in the present study and those from Ref.~\cite{Verkhovtsev_2022}, key characteristics and common features of the coating structure and distribution of water are analyzed for a broad range of coating densities.

The interconnection of structural and chemical parameters of the coating (composition, length, and surface density of ligand molecules) and NP size with the structural properties of water near the NP surface is an interesting and open research question, as these relations cannot be readily determined experimentally.
Computer simulations can provide necessary insights into these characteristics at the atomistic level, but they are limited to relatively small NP and ligand sizes.

A 3.7~nm gold NP has been selected as a case study in this paper.
NPs of that size, functionalised with PEG, were used in radiosensitisation experiments \cite{Sophie, GU2009196}. In those experiments, the particular system size was chosen for the relatively large surface area of gold NPs for efficient attachment of PEG molecules, low toxicity, and rapid cell penetration. The chosen system's size also falls within the range of ``ideal'' NP sizes of $\sim$1--50~nm in diameter to pass through the cell membrane without causing cellular damage \cite{IdealNPSize}. The 3.7-nm NPs are also large enough to be visible using transmission electron microscopy with a typical image resolution of $\sim$1~nm \cite{Sophie}.

Polymer physics can also provide insight into the structure of organic coatings of radiosensitising metal NPs. There is a significant body of research into the physics of polymers attached to flat and curved surfaces, stemming from the work of Flory \cite{Flory} and, in particular, de Gennes \cite{DeGennes}. The structural properties of polymer chains attached to flat surfaces is a well-understood problem, studied for the first time by Alexander \cite{Alexander} and de Gennes \cite{DeGennesFlat}. The extension of these models to account for polymers attached to curved surfaces was first made by Daoud and Cotton \cite{DaoudCotton} and further developed by Wijmans and Zhulina \cite{WijmansZhulina}, providing a theoretical description of the height of a polymer chain attached to a curved surface in the limit of both large and small curvatures (that is, metal cores with small and large radii, respectively).
The results of MD simulations presented in this study, in combination with the results of our recent study \cite{Verkhovtsev_2022} and those from the field of polymer physics \cite{Dukes, WijmansZhulina, PEO}, are used to calculate key structural parameters of the coatings of radiosensitising gold NPs and establish connections between the coating structure and distribution of water for different NP sizes as well as lengths and surface densities of coating molecules.
The results of the quantitative analysis of water distribution in the vicinity of coated metal NPs can subsequently be used in future studies to evaluate the radiosensitising effectiveness of a particular NP system based on the proximity of water to the NP metal core.

\section{Methodology}
\label{sec:Methodology}

This section briefly outlines the computational methodology for the creation and atomistic-level characterisation of 3.7-nm gold NPs coated with thiol-PEG-amine molecules and for complementary simulations of thiol-PEG-amine molecules attached to a flat gold surface. The computational protocol used to simulate coated 3.7-nm gold NPs solvated in water follows that outlined in the earlier study \cite{Verkhovtsev_2022}, including the creation of a metal core, coating of the metal core with thiol-PEG-amine molecules of different lengths, annealing of the coated systems in a vacuum, and, finally, solvation of systems in water. Modifications from the previously reported procedure are described in greater detail in the following sections.

In this study, classical MD simulations have been performed using the MBN Explorer software package \cite{MBNExplorer} for advanced multiscale modelling of complex molecular and nanoscale systems. MBN Studio \cite{MBNStudio}, a multitask toolkit and a dedicated graphical user interface for MBN Explorer, was used to construct the systems, prepare input files, and analyse simulation outputs.

\subsection{Thiol-PEG-amine coated gold nanoparticles}
\label{sec:Methodology_NPs}

Three PEG$_n$ molecules with the number of ethylene glycol monomers $n$ equal to 2, 8 and 14 have been considered in this study. The initial geometries of these molecules were taken from the LigandExpo database \cite{LigandExpo}.
Each PEG$_n$ molecule is functionalised with a thiol (--SH) group at one end and an amino (--NH$_2$) group at the other end, forming thiol-PEG$_n$-amine ($n=2, 8, 14$) molecules.
Chemical formulas and characteristics of the considered ligand molecules are listed in Table~\ref{table:PEG_Ligand_Specifications}.

\begin{table*}[t!]
\centering
\caption{Specification of three thiol-PEG$_n$-amine ($n = 2, 8, 14$) ligand molecules considered in this study. $N_{\text{atom}}$ is the number of atoms in each molecule; their chemical formulas and molecular weights are listed in the last two columns. The corresponding ID for the original molecules taken from the LigandExpo \cite{LigandExpo} database are given in brackets.}
\begin{tabular}{c|c|c|c|c}
        Short Ligand Name & Ligand Molecule & $N_{\text{atom}}$ & Formula & Mol. Weight (g/mol) \\ \hline
        PEG$_2$ (PGE) & S-(CH$_2$)$_2$-PEG$_2$-NH$_2$ & 24 & C$_6$H$_{14}$O$_2$NS & 162.3 \\
        PEG$_8$ (PE5) & S-(CH$_2$)$_2$-PEG$_8$-NH$_2$ & 66 & C$_{18}$H$_{38}$O$_8$NS & 428.6 \\
        PEG$_{14}$ (P4K) & S-(CH$_2$)$_2$-PEG$_{14}$-NH$_2$ & 107 & C$_{30}$H$_{62}$O$_{13}$NS & 677.9
    \end{tabular}
    \label{table:PEG_Ligand_Specifications}
\end{table*}

A spherical gold NP with a diameter of 3.7~nm, containing 1553 atoms, was created using the modeller tool of MBN Studio \cite{MBNStudio}. In experiments involving radiosensitizing gold NPs, the size of the NPs cannot be controlled precisely, and the typical size distribution of NPs is characterized by the standard deviation of about 1~nm. The constructed NPs thus represent a prototype of a statistically probable system synthesized in experiments involving radiosensitizing NPs. For large NPs the crystallographic planes within the gold core may impact the final coating morphology \cite{doi:10.1021/acsnano.8b09658}, however for the NP size considered in this work these effects are insignificant.

Thiol-PEG$_n$-amine molecules were then attached non-site-specifically to the NP surface as described in Ref.~\cite{Verkhovtsev_2022}. The number of attached molecules varied from 44 to 300, corresponding to the range of ligand surface densities $\sigma = 1.0 - 7.0$ molecules per nm$^{2}$. This range of ligand densities has typically been considered in experimental studies of coated gold NPs for biotechnology applications \cite{C6AN02206E}. The surface density of ligands $\sigma$ is defined as the number of ligands per unit surface area and is also commonly referred to as grafting density \cite{Benoit_2012_AnalChem.84.9238, Zhang_2015_DrugDeliv.22.182, PEO}. Seven values of ligand surface density have been considered for each ligand length. The total number of atoms in free gold NPs coated with thiol-PEG$_n$-amine ($n=2-14$) varied from $\sim$2600 atoms for the shortest ligand molecule and lowest surface density to $\sim$33,500 atoms for the longest ligand molecule and highest density cases, respectively.

In the computational protocol utilized in the present paper, the interaction between all gold atoms of the metal core (including atoms located on the NP surface and in its interior) has been modelled by means of the many-body Gupta potential \cite{GuptaPotential}. The covalent interaction between the gold atoms located on the metal core surface and sulphur atoms of the ligands is modeled using the Morse potential. Parameters of this interaction and non-covalent interactions between gold atoms and atoms of the ligands were previously validated through DFT calculations; see Ref.~\cite{Verkhovtsev_2022} for details. All interatomic interactions within the thiol-PEG-amine ligands and their interactions with water molecules after solvation were described using the CHARMM molecular mechanics force field \cite{CHARMM}. CHARMM parameters for the bonded and non-bonded interactions between atoms of the ligands and gold atoms are listed in Ref.~\cite{Verkhovtsev_2022}.

After the ligand attachment, the coated metal NPs were annealed (by heating the system from 0 to 600~K at a rate of 1~K/ps and then cooling the system back to 0~K at the same rate) before being introduced to a solvent.
The process of solvation of coated metal NPs in water follows the protocol outlined in Ref.~\cite{Verkhovtsev_2022}; the TIP3P water model \cite{TIP3P} was used to describe the solvent.
The number density of water molecules was set to 33.4~nm$^{-3}$, corresponding to the accepted number density of water molecules in a liquid medium at ambient conditions. All simulations were conducted for the canonical (NVT) ensemble of particles at 300 K and at constant pressure.
The geometry of solvated NPs was first optimised using the velocity quenching algorithm over 20,000 optimisation steps. Then the systems were equilibrated at 300~K using the Langevin thermostat for 5~ns.
The last 2~ns of each simulated trajectory were used for the structural characterisation analysis described in Section~\ref{sec:Results_PEG_Analysis}.

In the case of the gold NPs coated with thiol-PEG$_2$-amine ligands, a simulation box of $(10 \times 10 \times 10)$~nm$^3$ was used, and the total number of atoms in the system (including solvent) was equal to $\sim$106,000. For the case of thiol-PEG$_8$-amine ligands, the simulation box size was set to $(12 \times 12 \times 12)$~nm$^3$, with a characteristic system size of $\sim$182,000 atoms. Finally, for thiol-PEG$_{14}$-amine ligands, a simulation box of $(15 \times 15 \times 15)$~nm$^3$ was used, with the characteristic number of atoms in such systems equal to $\sim$360,000 atoms.

21 systems (7 ligand surface densities at 3 ligand lengths) have been modelled and characterised in this study.
The total simulation time was about 100,000 CPU hours.

\subsection{Thiol-PEG-amine coated gold surfaces}
\label{sec:Methodology_Surfaces}

Complementary simulations have also been performed to characterise the structure of thiol-PEG$_n$-amine ($n=2-14$) ligands attached to a flat gold substrate.
A gold slab with a size of $(4.08 \times 4.08 \times 1.5)$~nm$^3$ containing five layers of gold atoms was created from an ideal gold crystal by means of the modeller tool of MBN Studio \cite{MBNStudio}; the created substrate contained 1000 gold atoms. The simulations were carried out using periodic boundary conditions, such that the system is periodically translated in $xy$-plane.

As the next step, the gold substrate was coated with thiol-PEG$_n$-amine ($n = 2,8, 14$) molecules. Ligand molecules were evenly distributed above the substrate surface.
Three surface densities of ligands were chosen, namely 1.0, 2.3 and 7.0 molecules per nm$^2$. These particular values were chosen to describe the case of low, intermediate, and high surface densities corresponding to the range of values used for thiol-PEG$_n$-amine coated gold NPs. The number and spacing of thiol-PEG$_n$-amine molecules on top of the flat gold surface were dictated by the chosen surface density.

The process of attaching thiol-PEG$_n$-amine molecules to the substrate followed the same protocol outlined in Section~\ref{sec:Methodology_NPs} and in Ref.~\cite{Verkhovtsev_2022}. To avoid the translational motion of the whole system in the $z$-direction, the bottom layer of gold atoms was fixed during the process of ligand attachment. Following this, a series of constant-temperature MD simulations have been conducted to equilibrate the systems at 300~K for a period of 0.6~ns.

\begin{figure*}[t!]
\centering
\includegraphics[width=0.9\textwidth]{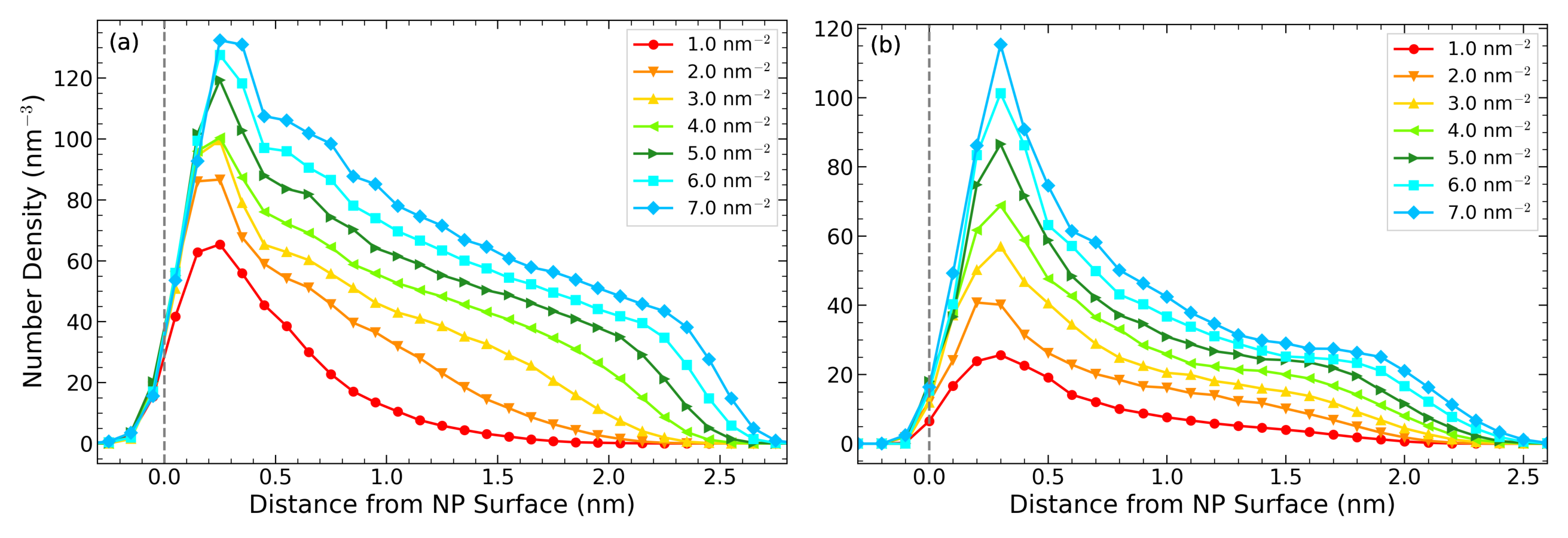}
\caption{Atomic number density $n(r)$ of thiol-PEG$_8$-amine coating for (a) 3.7~nm and (b) 1.6~nm diameter gold NPs solvated in water for ligand surface densities between 1.0 and 7.0 nm$^{-2}$. The corresponding number of molecules varies from 44 to 300 for a 3.7~nm NP and from 8 to 56 for a 1.6~nm NP. The atomic number densities are plotted as a function of the distance $r$ from the gold NP surface. The average position of the NP surface is indicated by the dashed grey line. Data for the 1.6~nm NP (panel~(b)) are taken from Ref.~\cite{Verkhovtsev_2022}.}
\label{fig:PEGPlots}
\end{figure*}

The equilibrated systems were solvated in water following the protocol outlined above. The simulation box size was set equal to $(4.08 \times 4.08 \times 6.0)$~nm$^3$ for thiol-PEG$_{2}$-amine and thiol-PEG$_{8}$-amine and $(4.08 \times 4.08 \times 7.0)$~nm$^3$ for thiol-PEG$_{14}$-amine. In the case of the highest ligand surface density of 7.0 nm$^{-2}$, the simulation box size in the $z$ direction was increased to 7.5~nm due to the elongation of the ligands as a result of electrostatic repulsion between ligand atoms. The solvated systems underwent energy minimisation to avoid any possible overlap between water and ligand molecules.

In order to create physically correct system states thermalised at a given temperature, geometry optimisation calculations were followed with two subsequent MD simulations. The first was a short 20~ps simulation at 300~K to pre-equilibrate the systems, followed by a longer 5~ns long MD simulation to thermalise the systems at 300~K.

\section{Results and discussion}
\label{sec:Results}

Structural analysis of the coatings of nanometer-sized gold NPs can provide valuable insight into the effect of coating structure on the distribution of water in the vicinity of the solvated NPs. Such an analysis is performed for NPs of two sizes (1.6 and 3.7~nm in diameter), different ligand lengths, and ligand surface densities. Based on the outcomes of this analysis, the relationship between the radial distribution of atoms of the coating and the distribution of water in the vicinity of the NPs is explored.

\subsection{Structural characterisation of thiol-PEG-amine coatings and their conformation states}
\label{sec:Results_PEG_Analysis}

Figure~\ref{fig:PEGPlots}(a) shows the atomic number density of a coating made of 44 to 300 thiol-PEG$_8$-amine molecules attached to 3.7~nm gold NPs (corresponding to the ligand surface density in the range from 1.0 to 7.0 nm$^{-2}$). The number density is plotted as a function of the average distance from the NP surface. The atomic number density has been calculated as the number of atoms of the coating in spherical bins of 0.1~nm thickness, divided by the volume of each bin.
The curves presented in Figure~\ref{fig:PEGPlots} and further below in this section were averaged over 50 independent frames taken from the final 2~ns long segments of the simulated trajectories.

Figure~\ref{fig:PEGPlots}(b) shows the results of a similar analysis \cite{Verkhovtsev_2022} conducted for 1.6~nm diameter gold NPs coated with thiol-PEG$_8$-amine molecules for the same range of ligand surface densities (the number of molecules varies from 8 to 56). Figure~\ref{fig:PEGPlots} illustrates that the profiles of the atomic number densities of the coating for the NPs of two different sizes are qualitatively similar. The maximum atomic number density of the coating is located at $\sim$0.35~nm from the NP surface in the case of a 3.7~nm NP and
at $\sim$0.3~nm from the surface for a 1.6~nm gold NP. The maximum atomic number density for the coating of the 3.7~nm NP is generally higher than that for the 1.6~nm NP, implying a dependence of the maximum number density of the coating on the NP radius.

\begin{figure}[t!]
\centering
\includegraphics[width=0.45\textwidth]{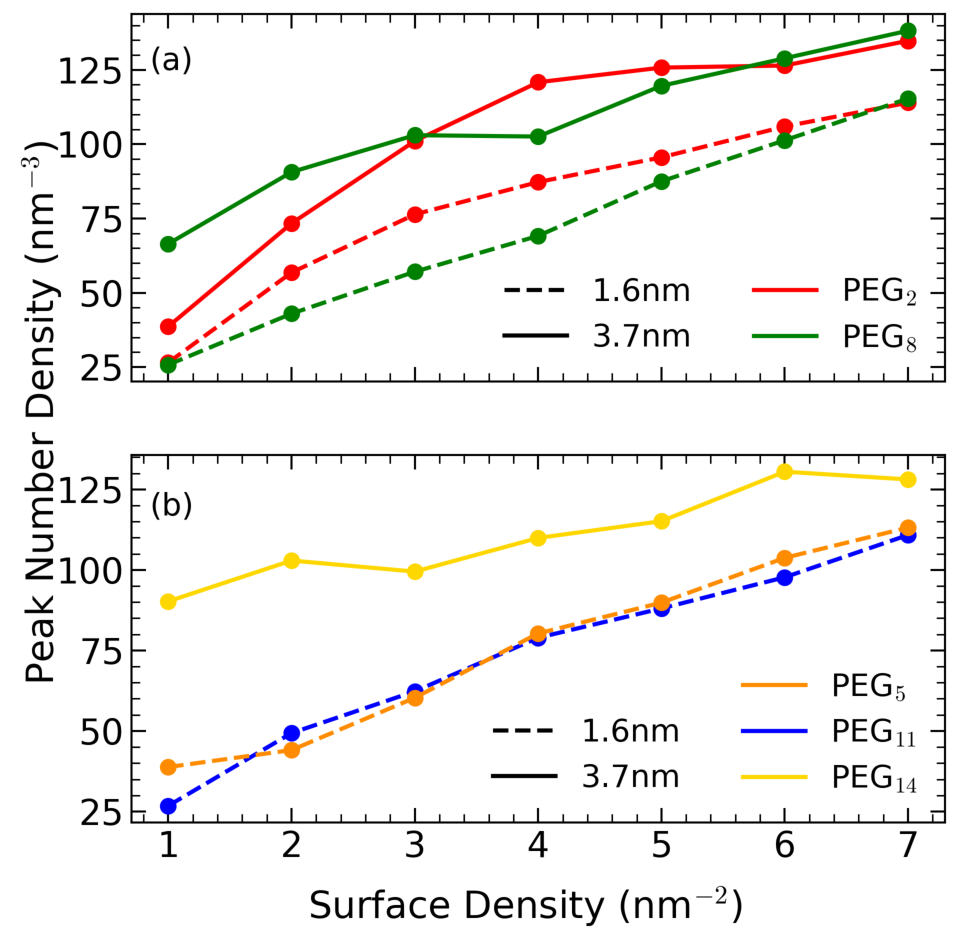}
\caption{Peak number density of thiol-PEG$_n$-amine ligands for $n=2,8$ (panel~(a)) and $n=5, 11, 14$ (panel~(b)), as a function of ligand surface density for 1.6~nm and 3.7~nm diameter NPs (dashed and solid curves, respectively). The peak values correspond to the maximum values of the number density profiles plotted in Fig.~\ref{fig:PEGPlots}.}
\label{fig:PeakValues}
\end{figure}

In both cases, a shoulder in the number density distributions is formed at distances $\sim 1.9 - 2.4$~nm from the surface of a 3.7~nm NP and at distances $\sim 1.6-2.2$~nm from the surface of a 1.6~nm NP. This shoulder appears at intermediate ligand surface densities of $\sigma \sim 4.0$~nm$^{-2}$ (172 and 32 ligands; see the light-green curves) but is most pronounced at densities larger than 5.0~nm$^{-2}$ (216 and 40 ligands; see the dark-green curves). The appearance of the shoulder is associated with a change between conformation states of the coating molecules from ``mushroom''-like structures to ``brush''-like structures \cite{Verkhovtsev_2022}. In polymer physics, conformation states describe the structures of polymer coating attached to surfaces. There are two main conformation states, namely ``mushroom'' and ``brush'' conformations \cite{DeGennesFlat, PolymerBrushes, Binder_2012_JPolymSciB.50.1515}. The ``mushroom'' conformation occurs when ligands are loosely packed together and do not interact with each other. In this case, the ligands are strongly bent. ``Mushroom'' confirmations typically occur at lower ligand surface densities. The ``brush'' conformation occurs at higher ligand densities when the ligands begin to interact with each other and repel each other. This interaction causes the ligands to straighten \cite{PolymerBrushes, DeGennes, DeGennesFlat, Dukes, WijmansZhulina}.

Figure~\ref{fig:PeakValues} shows the dependence of the maximum number density values of thiol-PEG$_n$-amine ligands on ligand surface density. In all the cases considered, a decrease in the derivative of this function is observed at characteristic surface densities, which approximately correspond to the surface densities of the transition between the ``mushroom''-like and ``brush''-like conformations.
The surface density value, at which the derivative decreases, varies with ligand length, occurring at lower surface densities for longer ligands.
The details of this transition are elaborated below in Section~\ref{sec:ConformationTransition}.
This transition is less pronounced for the smaller 1.6~nm diameter NP (dashed curves in Fig.~\ref{fig:PeakValues}), where a smooth increase in the peak number density is observed, while for the larger 3.7~nm diameter NP (solid curves), the change in the derivative is more evident.

\begin{figure}[t!]
\centering
\includegraphics[width=0.43\textwidth]{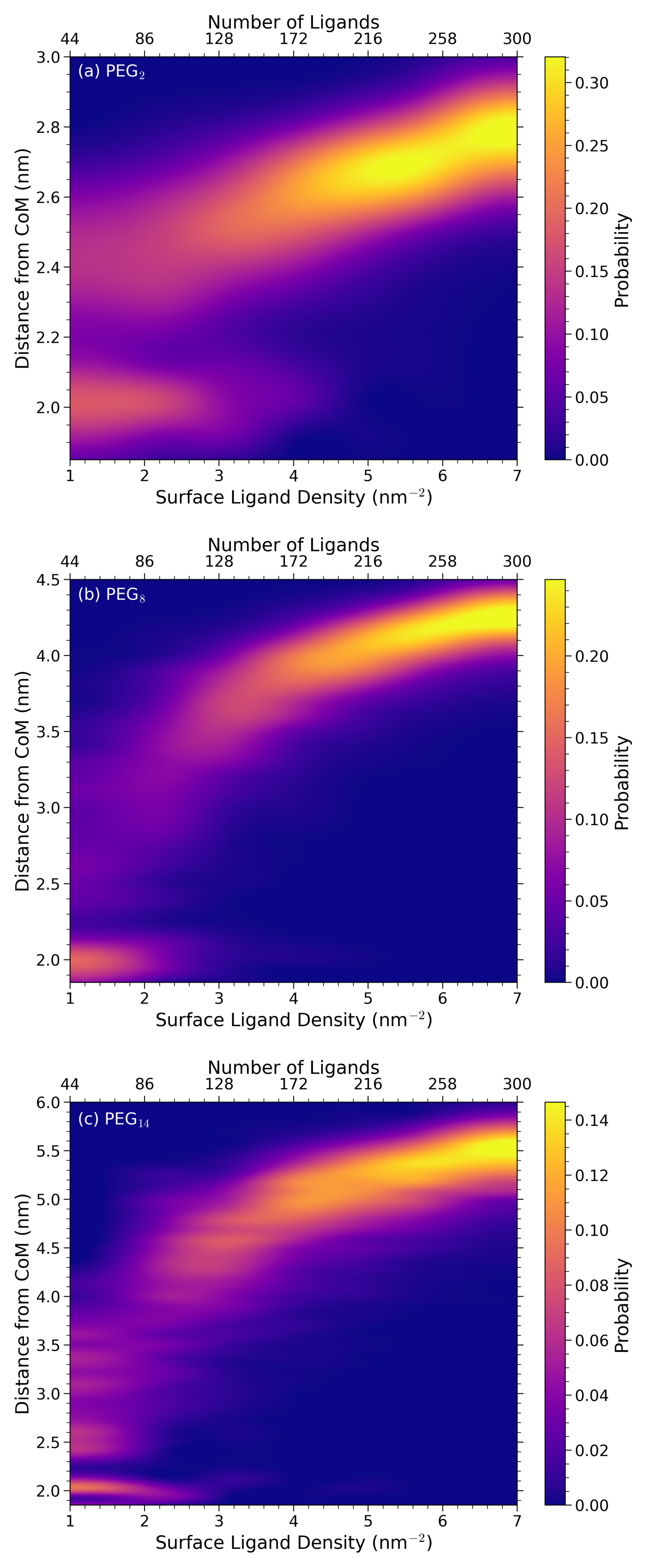}
\caption{Probability distribution of nitrogen atoms in thiol-PEG$_n$-amine coatings; (a) $n=2$, (b) $n=8$, (c) $n=14$. These are plotted as a function of surface density of ligands / the corresponding number of ligand molecules, and distance from NP center of mass (CoM). The color scale shows the probability of finding a nitrogen atom at a given distance from CoM. Note the different color scale in each panel.}
\label{fig:NitrogenHeatmaps}
\end{figure}

The atomic number density distributions plotted in Figure~\ref{fig:PEGPlots} can be understood further by analysing the transition from the ``mushroom''-like to ``brush''-like structures. Figure~\ref{fig:NitrogenHeatmaps} shows the probability distribution of finding nitrogen atoms of the terminal NH$_2$ groups of thiol-PEG$_n$-amine molecules ($n = 2,8,14$) at a given distance from the center of mass (CoM) of the NP. This is determined by summing the number of nitrogen atoms in 0.1~nm thick spherical bins and dividing it by the number of ligand molecules in the coating. The probability distribution is plotted as a function of ligand surface density and the corresponding number of ligand molecules.

As the nitrogen atoms are present in the terminal group of the ligands, their position is directly related to the conformation state seen \cite{Verkhovtsev_2022}. For lower density (fewer ligand) cases, an island of increased probability is seen close to the NP surface (located at the distance of $\sim$1.85~nm from CoM), corresponding to the ``mushroom''-like structures. For increasing surface density, the probability distribution shifts to favour ``brush''-like conformations. In all cases, a transition region is also seen, with a mixture of conformations being present for ligands surface densities up to $\sim 3.0-4.0$~nm$^{-2}$. At higher ligand densities, the conformation state of the coating is exclusively ``brush''-like. In Figure~\ref{fig:PEGPlots}(a), the shoulder begins to appear for surface densities of between 4.0 nm$^{-2}$ and 5.0 nm$^{-2}$ (corresponding to 172 and 216 ligands, respectively). As discussed below, the origin of the shoulder can be attributed to the position of the terminal NH$_2$ group. Thus, the presence of a mixture of ``mushroom''-like and ``brush''-like conformation states may suppress the formation of a shoulder at lower coating densities.

\subsubsection{Coating thickness}
\label{section:CoatingHeight}

\begin{figure}[t!]
\centering
\includegraphics[width=0.45\textwidth]{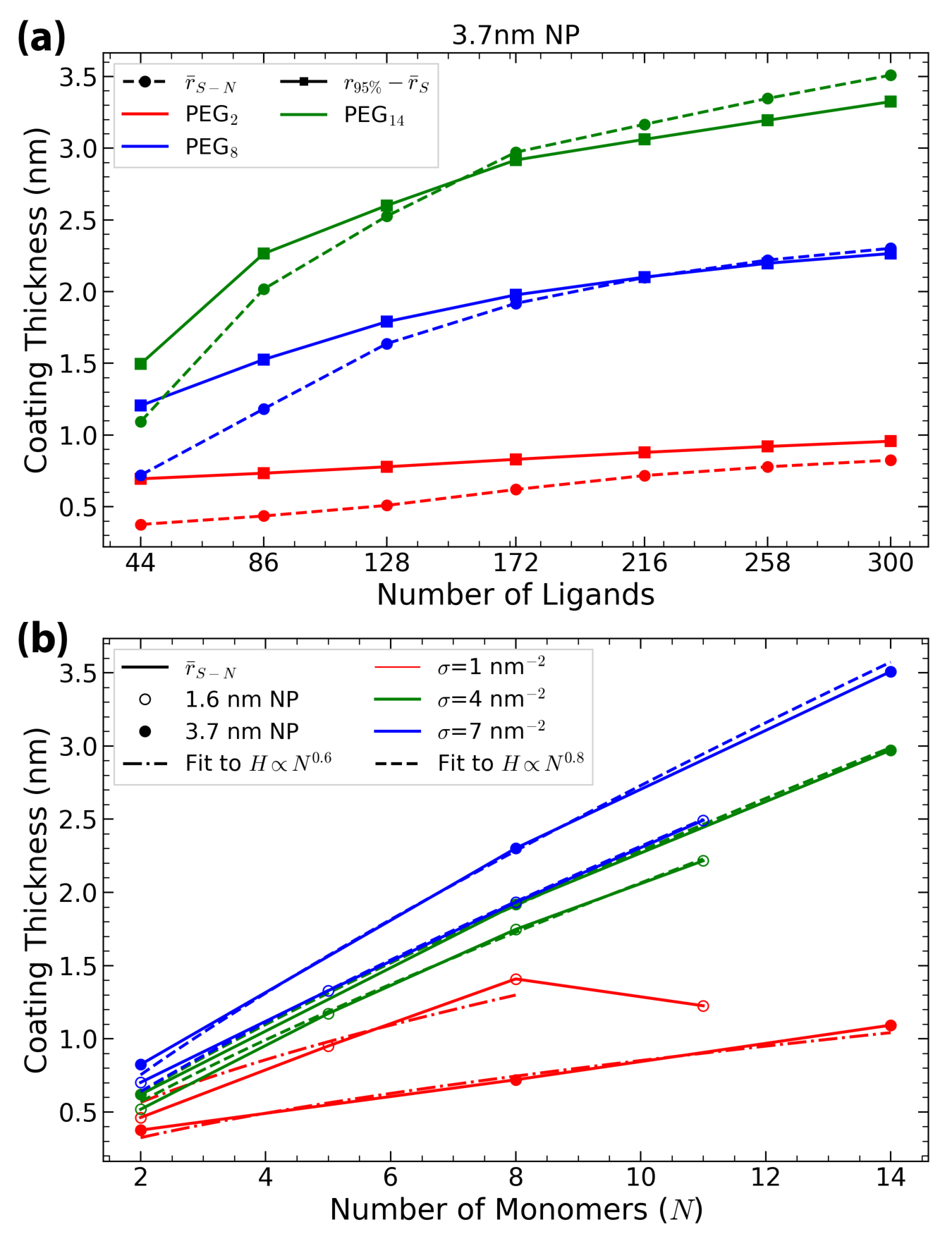}
\caption{Panel~\textbf{(a)}: The calculated thickness of thiol-PEG$_n$-amine ($n=2,8,14$) coatings as a function of the number of attached ligand molecules. Dashed lines with circles show the average distance between sulphur and nitrogen atoms, $\Bar{r}_{S-N}$. Solid lines with squares show the distance between the average position of sulphur atoms and the distance enclosing 95\% of all atoms of the coating, $r_{95\%}-\Bar{r}_S$. Panel~\textbf{(b)} shows the calculated thickness $L$ of thiol-PEG$_n$-amine coating as a function of the number of monomers $n$ for the 3.7~nm NP considered in the present study (solid symbols) and the 1.6~nm NP studied in Ref.~\cite{Verkhovtsev_2022} (open symbols). The calculated values are fitted to the $L \propto n^{0.6}$ and $L \propto n^{0.8}$ dependencies known from the polymer physics for the ``mushroom'' and ``brush'' polymer conformations, respectively \cite{Dukes}.}
    \label{fig:CoatingThickness}
\end{figure}

Different conformation states in thiol-PEG$_n$-amine ligands can also be identified through the calculation of the average thickness of the coating.
Figure~\ref{fig:CoatingThickness}(a) shows the thickness of the coatings made of thiol-PEG$_n$-amine ($n=2,8,14$) ligands attached to the surface of a 3.7~nm gold NP.
Dashed curves with circles show the average distance between sulphur and nitrogen atoms, $\Bar{r}_{S-N}$, and the solid curves with squares show the distance between the average position of sulphur atoms and the distance from CoM that encloses 95\% of all coating atoms, $r_{95\%} - \Bar{r}_S$.

At small numbers of attached ligands (i.e., at low surface densities), the distance $(r_{95\%} - \Bar{r}_S)$ exceeds the distance $\Bar{r}_{S-N}$ for all the cases considered, indicating that at lower surface densities, ligands are strongly bent, and thus they attain a ``mushroom''-like conformation. For the cases of thiol-PEG$_8$-amine and thiol-PEG$_{14}$-amine (blue and green curves, respectively) the $\Bar{r}_{S-N}$ distance exceeds the $(r_{95\%} - \Bar{r}_S)$ distance at higher ligand densities.
The S--N distance $\Bar{r}_{S-N}$ becomes larger than $(r_{95\%} - \Bar{r}_S)$ for the coatings made of $\sim$172 to 216 ligands. Such a transition from ``mushroom''-like to ``brush''-like structures is also seen in the nitrogen probability distributions shown in Figure~\ref{fig:NitrogenHeatmaps}, indicating that at higher ligand densities the coating takes on a ``brush''-like conformation. In the case of thiol-PEG$_2$-amine, the 95\% distance exceeds the S--N distance for all considered values of ligand densities, indicating that a purely ``brush''-like regime is not observed in this case. Rather, at lower ligand densities, the thiol-PEG$_2$-amine ligands have the ``mushroom''-like shapes, while at higher densities the coating represents a mixture of ``mushroom'' and ``brush''-like conformations.

The calculated values of coating thickness obtained from the MD simulations have been compared to the values predicted by polymer theory \cite{Dukes}. According to the latter, the thickness $L$ of a coating made of polymer molecules (containing $n$ repeating units) grafted to a curved surface scales as $L \propto n^{0.6}$ in the ``mushroom'' regime and $L \propto n^{0.8}$ in the ``brush'' regime \cite{Dukes}.
Figure~\ref{fig:CoatingThickness}(b) shows a fit of the calculated thiol-PEG$_n$-amine coating thickness as a function of the number of ethylene glycol units $n$ to these theoretical scaling laws. Solid symbols correspond to the case of a 3.7~nm gold NP considered in the present study, while open symbols denote the results from  Ref.~\cite{Verkhovtsev_2022} for a 1.6~nm gold NP.
Figure~\ref{fig:CoatingThickness}(b) demonstrates that the calculated coating thicknesses agree with the theoretical values, indicating that the results of present MD simulations resemble those expected from polymer physics.

\subsubsection{Radius of gyration}

Further insight into the structural properties and conformation of ligands in thiol-PEG-amine coatings can be obtained from the analysis of the radius of gyration of the ligands. The radius of gyration, $R_g$, is commonly used to characterise the size of a coiled polymer chain; it is defined as the root-mean-square distance from all the atoms of the molecule to its center of mass \cite{ElementsPolymerScienceBook}:
\begin{equation}
R_g = \left[ \frac{1}{M} \sum_i m_i \left( {\bf r}_i - {\bf r}_{\rm CoM}  \right)^2 \right]^{1/2} \ .
\label{eq:GyrationRadius}
\end{equation}
Here $M$ is the mass of the ligand molecule and ${\bf r}_{\rm CoM}$ is the position vector of its center of mass; $m_i$ and ${\bf r}_i$ are the mass and
the position vector of a constituent atom $i$, respectively.

\begin{figure}[tb!]
    \centering
    \includegraphics[width=0.48\textwidth]{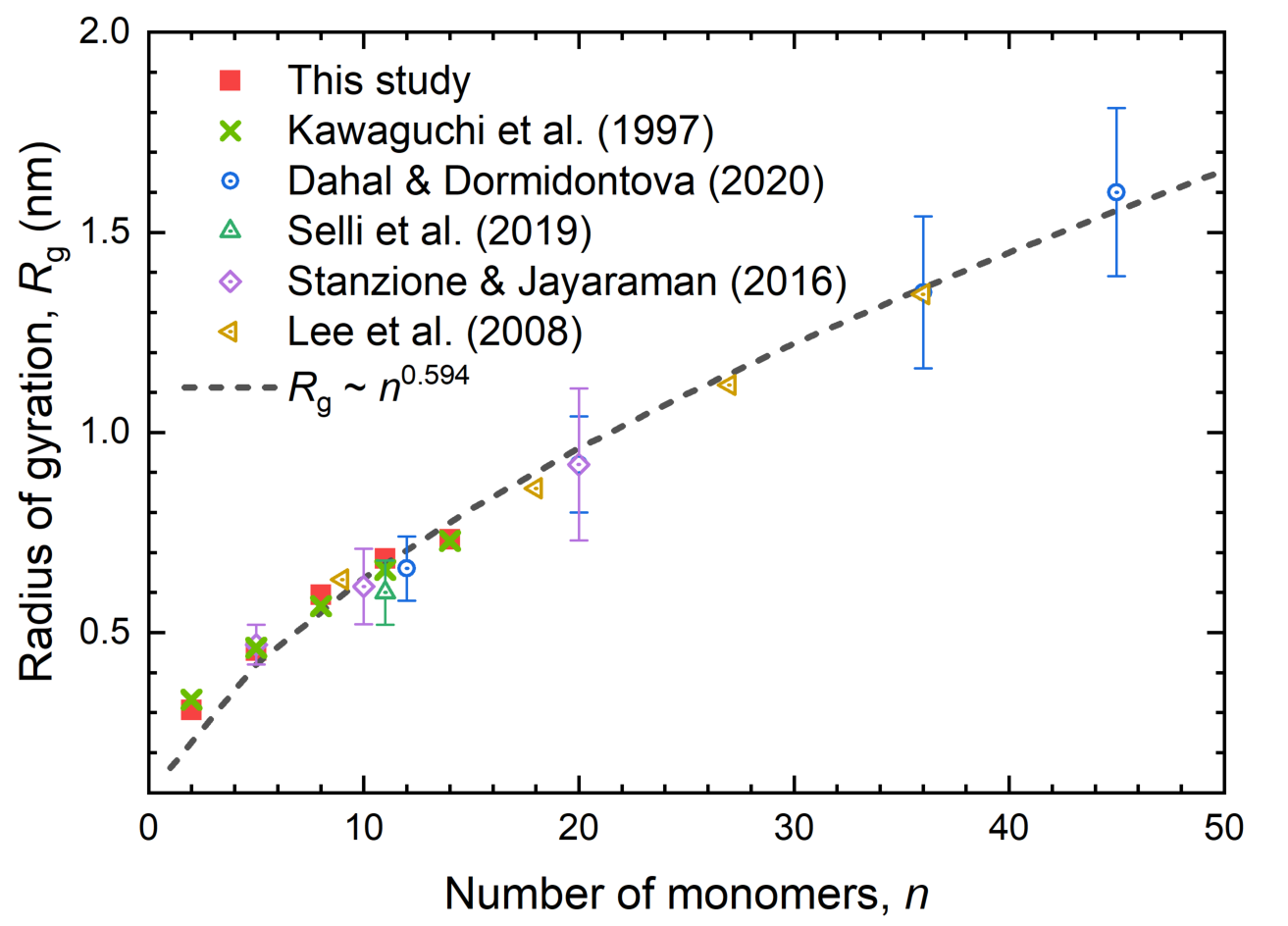}
    \caption{The radius of gyration $R_g$ of thiol-PEG$_n$-amine ligands calculated in the present study (red squares) and the radii of gyration for PEG$_n$ molecules (also known as polyethylene oxide, PEO) obtained from other MD simulations \cite{Selli_2019_JCollInterfSci.555.519, Stanzione_2016_JPCB.120.4160, Lee_2008_BiophysJ.95.1590, PEO} (open symbols) and experimental data obtained from static light scattering measurements \cite{Kawaguchi_1997_Polymer.38.2885} (green crosses).
    The dashed line shows a power law fit, $R_g\propto n^{\nu}$ with $\nu \approx 0.594$, for our results for the number of monomers $n=2-14$  and those reported in Ref.~\cite{PEO} for $n=12-45$.}
    \label{fig:GyrationRadiusComparison}
\end{figure}

In the present study, the radius of gyration for thiol-PEG$_n$-amine ($n = 2-14$) molecules has been determined through complementary MD simulations of single thiol-PEG$_n$-amine molecules solvated in water and thermalised at 300~K.
For each molecule, the radius of gyration was calculated according to Eq.~(\ref{eq:GyrationRadius}) by averaging over 50 frames of the simulated trajectory.
The calculated values of $R_g$ for thiol-PEG$_n$-amine molecules of different sizes
 are shown in Figure~\ref{fig:GyrationRadiusComparison} by red symbols.

The calculated values of $R_g$ for thiol-PEG$_n$-amine ($n=2-14$) molecules have been compared to the radii of gyration for PEG$_n$ molecules (also known as polyethylene oxide, PEO) reported in the literature. Figure~\ref{fig:GyrationRadiusComparison} presents the comparison of our data with those obtained from other MD simulations \cite{Selli_2019_JCollInterfSci.555.519, Stanzione_2016_JPCB.120.4160, Lee_2008_BiophysJ.95.1590, PEO} and experimental data obtained from static light scattering measurements \cite{Kawaguchi_1997_Polymer.38.2885}. In the latter case, the following scaling law between the radius of gyration (given in nanometers) and the molecular weight $M_{\textrm w}$ of a polymer molecule was proposed: $R_g = 0.0202 M_{\textrm w}^{0.55}$ \cite{Kawaguchi_1997_Polymer.38.2885}. The green crosses in Figure~\ref{fig:GyrationRadiusComparison} describe the values of $R_g$ calculated using the indicated scaling law for the molecular weights of thiol-PEG$_n$-amine ligands listed in Table~\ref{table:PEG_Ligand_Specifications} and taken from Ref.~\cite{Verkhovtsev_2022}. Overall, the $R_g$ values determined in the present study agree nicely with the results reported in the literature.

Open blue symbols in Figure~\ref{fig:GyrationRadiusComparison} show the results of a recent study \cite{PEO} where the $R_g$ values were determined for different PEO$_n$ ($n = 12,20,36,45$) molecules on the basis of MD simulations.
The figure shows the dependencies of $R_g$ on the number of monomers $n$, obtained in the present study in the range of $n=2-14$ and in Ref.~\cite{PEO} for $n=12-45$, follow a similar trend.
Fitting of the combination of the present $R_g$ values and those from Ref.~\cite{PEO} yields a power law dependence $R_g\propto n^{\nu}$, where $\nu \approx 0.594$ (see the dashed curve in Figure~\ref{fig:GyrationRadiusComparison}).
This result is in agreement with the power law dependence of $R_g \propto n^{0.549}$ determined through MD simulations for polyacrylic acid (PAA) polymers \cite{PAA}.
The exponent in these equations is known as the Flory Exponent $\nu$ and defines the type of polymer chain described. In the case of a real polymer chain, $\nu=3/5$ \cite{PolymerSolutionsBook}, which is close to the value $\nu = 0.594$ determined in the present study.

\subsubsection{Ligand density for the mushroom to brush transition} \label{sec:ConformationTransition}

Using the values of $R_g$ determined in the previous section, one can evaluate the ligand surface density of thiol-PEG$_n$-amine molecules at which the transition from ``mushroom''-like to ``brush''-like structures occurs for the ligands of different sizes.
As follows from the polymer theory \cite{Dukes}, at low ligand surface density, $\sigma \lesssim 1/R_g^2$, the radius of gyration of the ligands does not exceed the spacing between the neighboring ligands attached to a surface. In this case, individual ligands do not interact with adjacent ligands, thus acquiring ``mushroom''-like shapes.
Consequently, the coating thickness in this regime is equal to $L \approx 2R_g$ \cite{DeGennesFlat, Dukes, Kim}. When the ``mushroom''-like structures begin to overlap (which happens at surface densities $\sigma \gtrsim 1/R_g^2$), a ``mushroom''-to-``brush'' transition occurs, and the ligands acquire ``brush''-like shapes.

\begin{figure}[t!]
    \centering
    \includegraphics[width=0.46\textwidth]{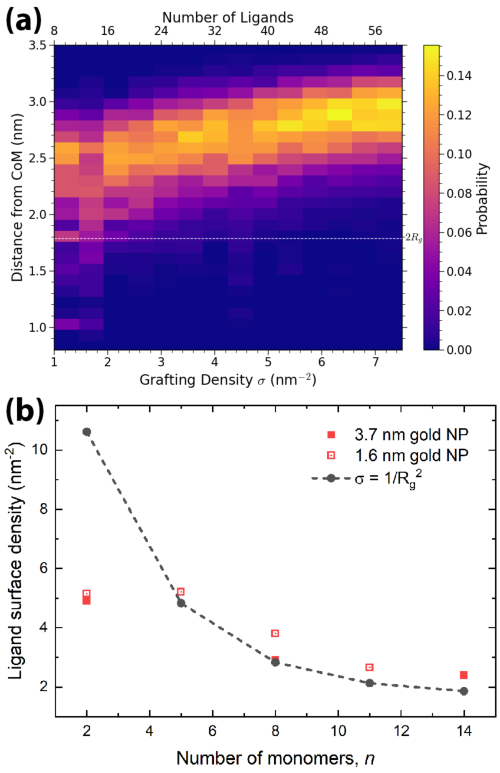}
    \caption{Panel~\textbf{(a)}: Plot of the probability distribution of nitrogen atoms located at a given distance from the NP CoM for various ligand surface densities / number of thiol-PEG$_8$-amine ligands. Twice the gyration radius is indicated by a dashed line to show the approximate ``mushroom''-like structure height.
    Panel~\textbf{(b)}: The estimated transition ligand density for 1.6 and 3.7~nm NPs coated with thiol-PEG$_n$-amine ligands of different lengths (closed and open red symbols). The dashed curve with black symbols shows the $\sigma = 1/R_g^2$ dependence, where the radius of gyration $R_g$ for the ligands containing $n$ monomer units has been calculated as plotted in Fig.~\ref{fig:GyrationRadiusComparison}.}
    \label{fig:CriticalGraftingDenmsityExample}
\end{figure}

The transition ligand density between the ``mushroom'' and ``brush'' regimes has been estimated from the results of performed MD simulations by analysing the probability distributions of the terminal nitrogen atoms within the NP coatings, similar to the results plotted in Fig.~\ref{fig:NitrogenHeatmaps}.
In this analysis, we have summed up the probabilities of finding nitrogen atoms at distances $r < 2R_g$ and $r > 2R_g$ from the NP metal surface for various surface densities of ligands.
In the ``mushroom'' regime, the height of the polymer coating is approximately equal to twice the radius of gyration, $h\approx 2R_{g}$ \cite{Kim}. Hence, by taking the sum of the nitrogen atom probability in the range of radial distances $R_c \le r \le R_c + 2R_g$ (where $R_c$ is the radius of the metal NP core), the transition ligand density has been evaluated for each particular case.
It has been assumed that the transition density corresponds to the density at which less than 10\% of nitrogen atoms are located at distances $r < 2R_g$ and more than 90\% of nitrogen atoms are located at $r > 2R_g$ from the metal core surface.
Figure~\ref{fig:CriticalGraftingDenmsityExample}(a) illustrates this method for the case of the 1.6~nm gold NP coated with thiol-PEG$_8$-amine at different surface densities.
The transition ligand density calculated this way for 1.6 and 3.7~nm NPs coated with thiol-PEG$_n$-amine ligands of different lengths is plotted in Figure~\ref{fig:CriticalGraftingDenmsityExample}(b). The discrepancy between the data points of the estimated transition surface density and the $1/R_g^2$ dependence (see the dashed curve) can be attributed to the nature of this transition. As shown in Fig.~\ref{fig:NitrogenHeatmaps}, at lower surface densities, the coating structure is not exclusively ``mushroom'' but rather a mixture of both ``mushroom'' and ``brush''-like conformations.

\subsection{Model for the number density of thiol-PEG-amine coating}

The results of the present MD simulations for a 3.7~nm gold NP and the earlier results for a 1.6~nm gold NP \cite{Verkhovtsev_2022} indicate that radial dependence of the atomic number density of the thiol-PEG$_n$-amine coatings has several common features for different NP sizes, ligand lengths, and surface densities of ligands. As such, one can develop a model that describes the number density profile of the coating for different parameters of the coating.
The considerations presented below in this section concern the range of intermediate and high ligand surface densities ($\sigma \gtrsim 4.0$ nm$^{-2}$) when the ligand molecules attain ``brush''-like shapes. 

\begin{figure}[tb!]
    \centering
    \includegraphics[width=0.45\textwidth]{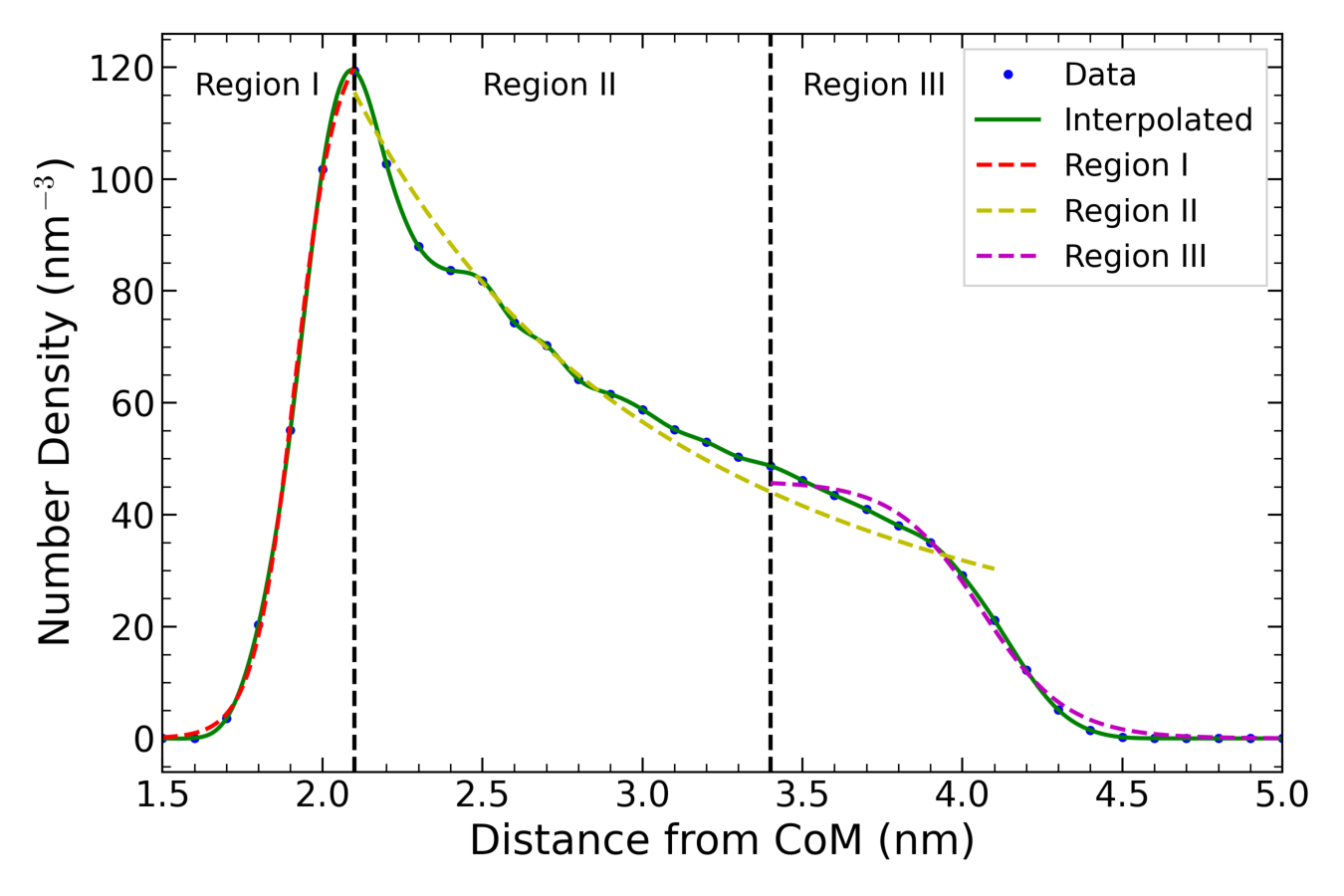}
    \caption{Radial distribution of atomic number density of a thiol-PEG$_8$-amine coating made of 216 ligands (surface density $\sigma = 5.0$ nm$^{-2}$) attached to a 3.7~nm NP. Symbols show the results of present MD simulations. Coloured dashed curves indicate three characteristic regions in the radial distribution of atomic number density of a thiol-PEG-amine coating, see Eqs.~(\ref{equation:RegionI})--(\ref{equation:RegionIII}). Solid green curve shows a continuous function combining three indicated regions using Eqs.~(\ref{equation:LinkFunction})--(\ref{equation:FullEquationSimple}); see the main text for details.}
    \label{fig:RegionsPlot}
\end{figure}

The atomic number density profiles for different thiol-PEG$_n$-amine coatings shown in Figure~\ref{fig:PEGPlots} can be split into three regions, which are illustrated in Figure~\ref{fig:RegionsPlot} for the case of 216 thiol-PEG$_8$-amine ligands (surface density $\sigma = 5.0$ nm$^{-2}$) attached to a 3.7~nm gold NP.

\begin{figure}[t!]
    \centering
    \includegraphics[width=0.45\textwidth]{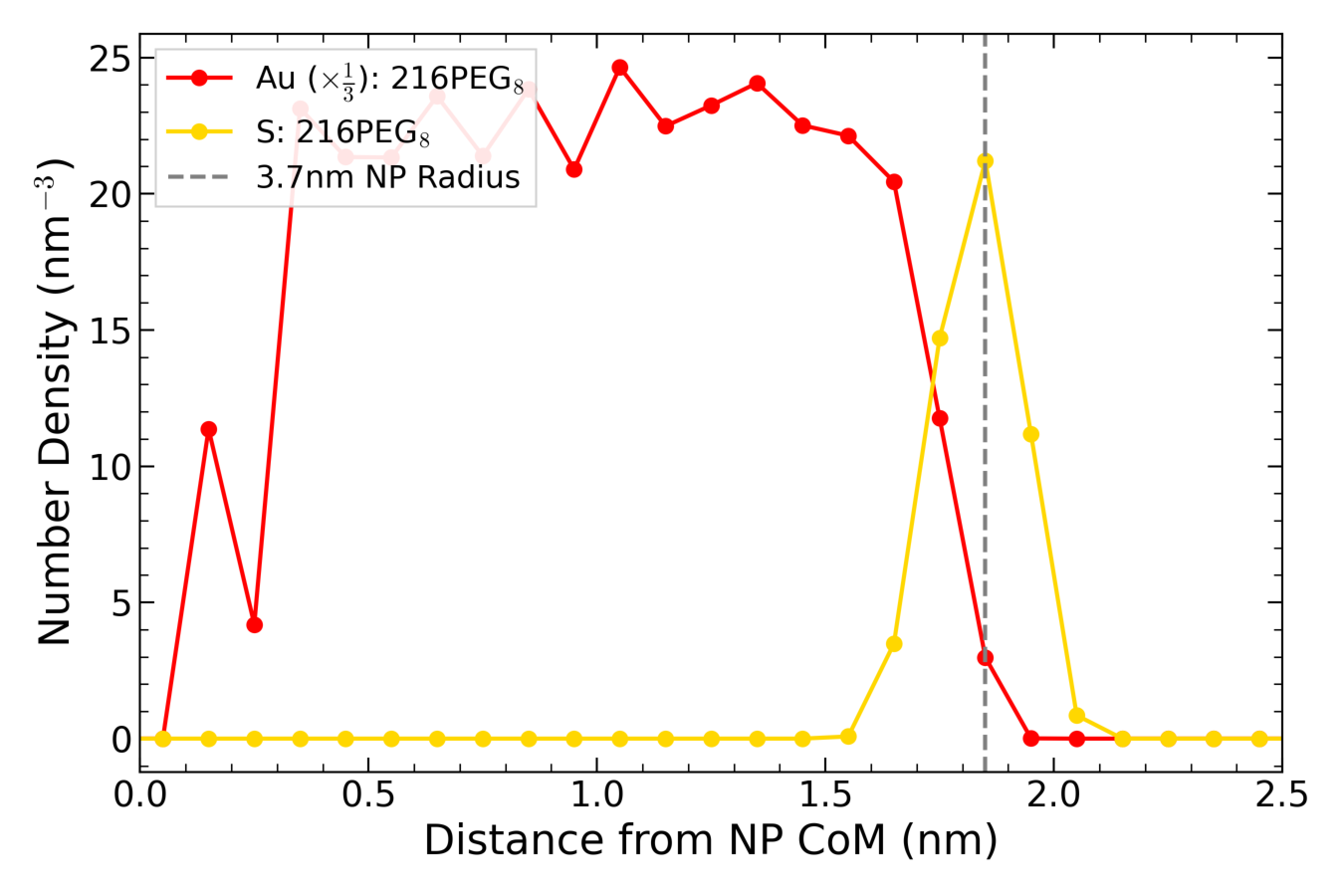}
    \caption{Number densities of sulphur atoms of thiol-PEG-amine ligands and gold atoms of the NP core as functions of the distance from the CoM.
    The figure shows an exemplar case of 216 thiol-PEG$_8$-amine ligands (corresponding to the surface density $\sigma = 5.0$ nm$^{-2}$) coating a 3.7 nm NP. The number density of gold has been scaled by a factor of $\frac{1}{3}$.}
    \label{fig:GoldPlots}
\end{figure}

The first region proximal to the NP surface (denoted as Region~I; see the dashed red curve in Fig.~\ref{fig:RegionsPlot}) is characterised by a rapid increase in the atomic number density of coating atoms up to the maximum value.
Figure~\ref{fig:PEGPlots} shows that for a particular NP size, the gradient of this rise is approximately the same for different values of the ligand surface density. As such, the radial distribution of number density in this region is determined by the distribution of ligands over the gold NP surface.
Figure~\ref{fig:GoldPlots} shows the radial distribution of the number density of sulphur atoms in the thiol group of thiol-PEG-amine ligands and the number density of gold atoms in the metal core.
The distribution of sulphur atoms is centred around the average radius of the NP, $R_c = 1.85$~nm, while the number density of gold atoms starts to decrease at smaller distances (of $\sim$1.6~nm) from the CoM.
This indicates that the metal NP core, after the annealing in a vacuum followed by solvation in water and thermalisation at 300~K, becomes non-spherical. Indeed, the NP surface has many facets, and sulphur ligands are bound across the gold surface based on the NP shape. Therefore, the distribution of ligands over the NP surface is directly linked to the distribution of the surface gold atoms of the NP core. As such, it is expected that atomic number density profiles for Region~I will depend on the NP size.

The maximum number density for atoms of the coating can be determined using the following geometric model outlined in Ref.~\cite{Verkhovtsev_2022}. The volume occupied by each repeating structural unit in thiol-PEG-amine ligands can be approximated as a cylinder with height $a$ (that is a characteristic monomer length) and base radius $r_0$. Considering the NP core as a sphere of radius $R_c$, the number of ligands that can be distributed across the surface, $N_{\text{lig}}$, is determined through the ratio of the NP core surface area, $A_{\text{core}}$, to the cross-sectional area of a single ligand, $A_{\text{lig}}$,
\begin{equation}
N_{\text{lig}} = \frac{A_{\text{core}}}{A_{\text{lig}}} = 4\left(\frac{R_c}{r_0}\right)^2 \ .
\end{equation}
The number density of atoms in the coating is given by the number of atoms in spherical bins of constant thickness $\Delta r$ divided by the volume of each bin.

There are, on average, eight atoms per monomer in a thiol-PEG-amine ligand. Taking the bin thickness equal to the characteristic linear size of a single monomer, $\Delta r = a$, one obtains the following expression for the maximum number density of atoms in the coating:
\begin{equation}
n_{\text{max}} = 24\frac{R_c^2}{\pi r_0^2\left[\left(R_c+a\right)^3 - R_c^3\right]} \ .
\label{equation:MaxNumberDensity}
\end{equation}
Using a monomer length of $a\approx 0.25$ nm and cross-sectional radius $r_0\approx 0.26$ nm, one obtains the maximum number density $n_{\text{max}} \approx 132$ nm$^{-3}$ for a 3.7~nm NP and $n_{\text{max}} \approx 112$ nm$^{-3}$ for a 1.6 nm NP, which agree with the data plotted in Fig.~\ref{fig:PEGPlots}.

For Region~I, the dependence of atomic number density on the radial distance from the CoM can be described as part of a sigmoid function,
\begin{equation}
n_1(r) = \frac{L_1}{1+e^{-k_1(r-r_{1}^{\prime})}} \ ,
\label{equation:RegionI}
\end{equation}
where $L_1$ defines the maximum value of the sigmoid, equal to the maximum number density for a given ligand surface density.
$k_1$ defines the gradient of the sigmoid and is, therefore, a function of the distribution of sulphur and gold atoms (see Fig.~\ref{fig:GoldPlots}) that depend on the NP radius but not on the number of monomers. $r_{1}^{\prime}$ defines the midpoint of the sigmoid.
The analysis performed in this study suggests that all parameters that define Region~I are dependent on both the NP size and surface density of ligands, but
not the number of monomers in ligand molecules.

\begin{figure}[tb!]
    \centering
    \includegraphics[width=0.48\textwidth]{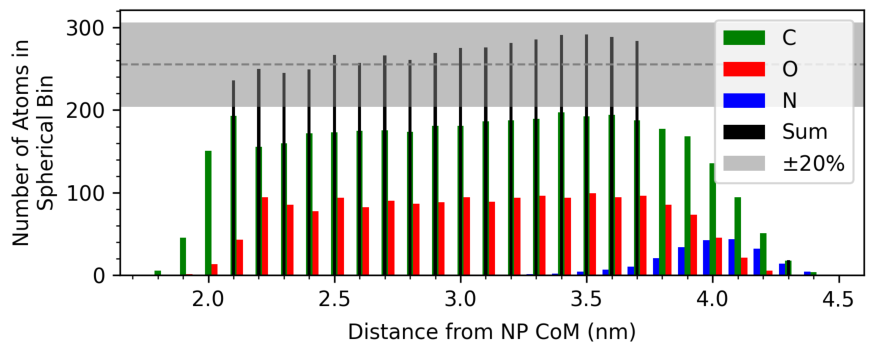}
    \caption{The number of carbon, oxygen, and nitrogen atoms of a thiol-PEG-amine coating in spherical bins of 0.1~nm thickness as a function of the distance from the NP CoM. The particular distribution represents the case of 216 thiol-PEG$_8$-amine ligands (surface density $\sigma = 5.0$~nm$^{-2}$) attached to a 3.7~nm gold NP. The sum of the carbon and oxygen atoms is shown as black bars, with the average value indicated by the grey dashed line; the shaded area shows a 20\% variation from the average value.}
    \label{fig:AtomNumbers}
\end{figure}

In Region~II (see the yellow dashed curve in Fig.~\ref{fig:RegionsPlot}), the coating number density has a power law dependence on the radial distance from CoM,
\begin{equation}
n_2(r) = \frac{L_2}{r^{\alpha}} \ .
\label{equation:RegionII}
\end{equation}
Here $L_2$ is a scaling parameter defined by the NP size and ligand surface density, and $\alpha \approx 2$.
The $n_2(r) \propto 1/r^{2}$ dependence stems from the distribution of the carbon and oxygen atoms that make up the backbone of the thiol-PEG-amine ligands \cite{Verkhovtsev_2022}. Figure~\ref{fig:AtomNumbers} shows the number of carbon, oxygen, and nitrogen atoms in spherical bins of 0.1~nm thickness in the coating made of 216 thiol-PEG$_8$-amine ligands (surface density $\sigma = 5.0$ nm$^{-2}$) attached to a 3.7 nm NP. The distribution of carbon and oxygen atoms in such a coating is approximately uniform (with the relative variation of $\sim$20\%) at distances from $\sim$2.0 to 4.0~nm from the CoM, while the surface area of each spherical bin increases as $r^2$. The approximately constant number of atoms in each bin results in the $1/r^{2}$ decrease in the number density.

The atomic number density for all ligand lengths and surface densities considered in the present study and Ref.~\cite{Verkhovtsev_2022} was fitted with the power law function, Eq.~(\ref{equation:RegionII}), in the range of distances from CoM where the radial distribution of carbon and oxygen atoms is approximately uniform. The resulting
values of the exponent $\alpha$ are $1.97 \pm 0.25$ for the 1.6~nm gold NP and $\alpha = 2.03 \pm 0.35$ for the 3.7~nm NP. This result indicates that the $n_2(r) \propto 1/r^{2}$ dependence is a general structural feature of the studied polymer coatings, independent of the NP size and length of ligand molecules.

Region~III (see the purple dashed curve in Fig.~\ref{fig:RegionsPlot}) is characterised by the appearance of a shoulder in the radial distribution of the coating number density. This shoulder is located at the position of the terminal NH$_2$ group of the thiol-PEG-amine ligands (see the distribution of nitrogen atoms in a thiol-PEG$_8$-amine coating made of 216 ligands in Figure~\ref{fig:AtomNumbers}). The distribution of nitrogen atoms shown in Figure~\ref{fig:AtomNumbers} is centred around a distance from the CoM of $\sim$4.0 nm, which corresponds to the position of the shoulder in the number density profile shown in Fig.~\ref{fig:PEGPlots}(a).

This shoulder appears at ligand surface densities $\sigma \gtrsim 4.0$~nm$^{-2}$ and, therefore, is also related to the conformation state of ligand molecules within the coating. Higher density coating results in a ``brush'' conformation due to electrostatic repulsion between densely packed ligands, resulting in the position of the terminal NH$_2$ groups being extended to larger distances from the CoM (see Fig.~\ref{fig:NitrogenHeatmaps}). As demonstrated in Figure~\ref{fig:PEGPlots}, the position of the shoulder is shifted to larger radial distances from CoM as the ligand surface density increases.
Note that no such shoulder has been observed in the radial distribution of coating number density at lower ligand densities $\sigma \lesssim 4.0$~nm$^{-2}$ (see Fig.~\ref{fig:PEGPlots}). This behaviour is explained by the ``mushroom''-like conformation of ligand molecules at low ligand surface densities so that the terminal NH$_2$ groups are scattered more uniformly over a broader range of distances from CoM.

The formation of a shoulder in the number density distribution can be understood from the random walk of the free end of a ligand molecule.
For ligands attached to a curved metal NP surface, each ligand molecule can be described as a flexible polymer with one of its ends being fixed at the tip
of a cone-shaped area.

The problem of a polymer constrained inside a cone was studied theoretically using the random walk theory \cite{Guttman_1980_PolymerCone}.
In the cited study, the probability density for a polymer chain end to be located at a given point in space after a given number of random steps has been determined.
It was found that the probability density as a function of the radial coordinate is described by a Gaussian distribution \cite{Guttman_1980_PolymerCone}.
Based on these considerations, the radial dependence of atomic number density for Region~III can also be characterised by a part of a sigmoid function:
\begin{equation}
n_3(r) = \frac{L_3}{1+e^{k_3(r - r_{3}^{\prime})}} \ .
\label{equation:RegionIII}
\end{equation}
Here $L_3$ is the maximum value of the sigmoid function (equal to the lowest number density for Region~II); $k_3$ depends on the width of the distribution of nitrogen atoms and, therefore, it becomes steeper with increasing surface density; $r_{3}^{\prime}$ is the sigmoid's midpoint.
In contrast to Regions~I and II, the parameters characterising Region~III depend on the NP size, ligand surface density, and the number of monomers in a ligand molecule.

The three regions of the number density profile introduced above can be combined into a piecewise function to describe the radial dependence of the atomic number density of ligands in terms of several geometrical parameters of the system:
\begin{equation}
    n(r) =
    \begin{cases}
        \frac{L_1}{1+e^{-k_1(r-r_{1}^{\prime})}}  & , \  r<R_1 \\
        \;\;\;\;\;\; \frac{L_2}{r^2}   & , \ R_1 < r < R_2 \\
        \frac{L_3}{1+e^{k_3(r-r_{3}^{\prime})}}   & , \  r>R_2
    \end{cases} \ .
    \label{equation:PiecewiseFunction}
\end{equation}
Here $R_1$ and $R_2$ are the radial distances from CoM defining the borders of the regions (see black dashed lines in Fig.~\ref{fig:RegionsPlot}), which depend on the NP radius; apart from that, $R_2$ is dependent on the number of ligand monomers.

The continuity of the resulting function can be achieved by combining the piecewise function, Eq.~(\ref{equation:PiecewiseFunction}), with a set of linking functions $f_L$ which take the form \cite{ContinuousFunction}
\begin{equation}
    f_L(x) = 1 - \left(\frac{1}{1+\left(\frac{x}{a}\right)^b}\right) \ .
    \label{equation:LinkFunction}
\end{equation}
Here $a$ is the point where the two functions intersect, and $b$ is a parameter that makes the joining of the linked functions smoother. The resultant continuous function is then given by
\begin{equation}
f(x) = f_1(1-f_L) + f_2f_L \ ,
\end{equation}
where $f_1$ and $f_2$ are the two parts of the piecewise function, Eq.~(\ref{equation:PiecewiseFunction}), that are being linked. To link a third part of the function, one uses another linking function along with the output of the first two linked functions.

Linking all parts of the piecewise function, Eq.~(\ref{equation:PiecewiseFunction}), yields the following continuous function for the radial dependence of the atomic number of density of the coating:
\begin{equation}
    \begin{aligned}
            n(r) = [n_1(r)(1-f_{L_1}) \\
            + n_2(r)f_{L_1}](1-f_{L_2}) + n_3(r)f_{L_2} \ ,
        \label{equation:FullEquationSimple}
    \end{aligned}
\end{equation}
where $n_1(r)$, $n_2(r)$ and $n_3(r)$ are given by Eqs. (\ref{equation:RegionI}), (\ref{equation:RegionII}) and (\ref{equation:RegionIII}), respectively.

\subsection{Structural characterisation of thiol-PEG-amine ligands grafted to a flat gold surface}

\begin{figure}[t!]
    \centering
    \includegraphics[width=0.49\textwidth]{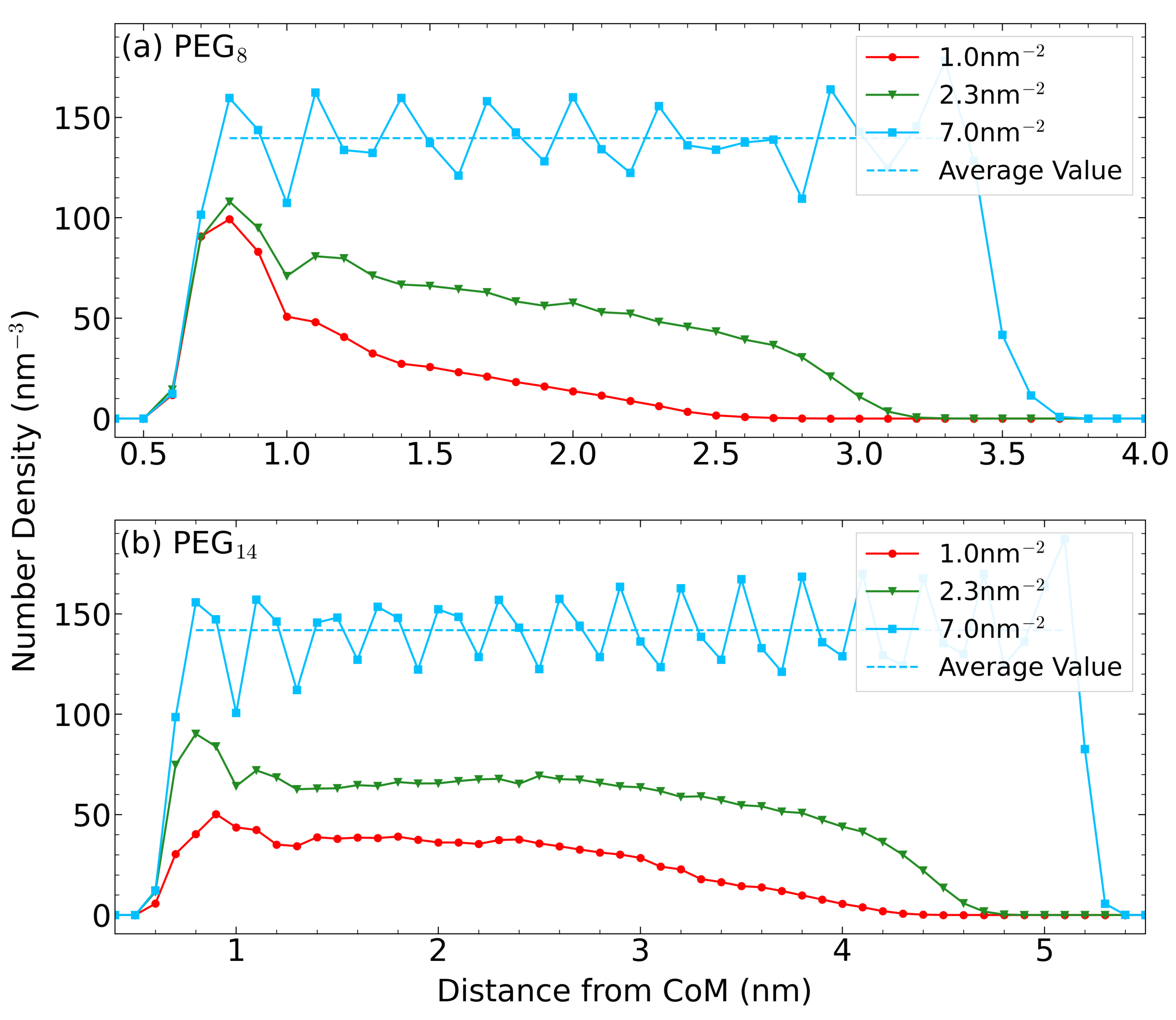}
    \caption{Atomic number density $n(r)$ of all atoms in thiol-PEG$_8$-amine and thiol-PEG$_{14}$-amine coatings grafted to a flat gold substrate as a function of the distance $r$ from the gold surface. Three values of ligand surface density, $\sigma = 1.0, 2.3$ and 7.0~nm$^{-2}$ have been considered for each ligand length.}
    \label{fig:FlatSurfaceNumberDensityPlot}
\end{figure}

It is also interesting to investigate how the structural properties of thiol-PEG-amine coatings on top of a gold surface depend on the system's geometry. For that, a similar analysis to the one presented above in Section~\ref{sec:Results_PEG_Analysis} has been performed for the thiol-PEG-amine coatings attached to a flat gold surface.
The process for calculating the atomic number density of the coating on a flat substrate is much the same as that for coated gold NPs. The number of coating atoms in slabs of 0.1~nm thickness, starting from the gold surface, was determined and then divided by the volume of each slab. Figure~\ref{fig:FlatSurfaceNumberDensityPlot} illustrates the number density of thiol-PEG$_8$-amine and thiol-PEG$_{14}$-amine coatings as a function of the distance from the substrate for three values of ligand surface density, $\sigma = 1.0, 2.3$ and 7.0~nm$^{-2}$.

Figure~\ref{fig:FlatSurfaceNumberDensityPlot} displays several notable features. First is the presence of a series of peaks in the number density distributions for ligand surface densities of 7.0 nm$^{-2}$ (blue curves). Each of these peaks corresponds to an individual ethylene glycol monomer unit, implying that no water has penetrated between the ligands. 
In the lower density cases, $\sigma = 1.0$ and 2.3~nm$^{-2}$, the number density profiles resemble those shown in Figure~\ref{fig:PEGPlots} for similar ligand surface densities.

It should be noted that the in the case of a flat surface, there is no power law dependence of $n(r)$ on $r$, as described by Region~II for the case of a metal NP. As such, the power $\alpha$ characterising the power law behaviour in Region~II should be related to the geometry of the metal surface, $\alpha \approx 2$ for the spherical geometry and $\alpha \approx 0$ for a flat geometry.

\subsection{Distribution of water around coated NPs}

\begin{figure}[t!]
    \centering
    \includegraphics[width=0.45\textwidth]{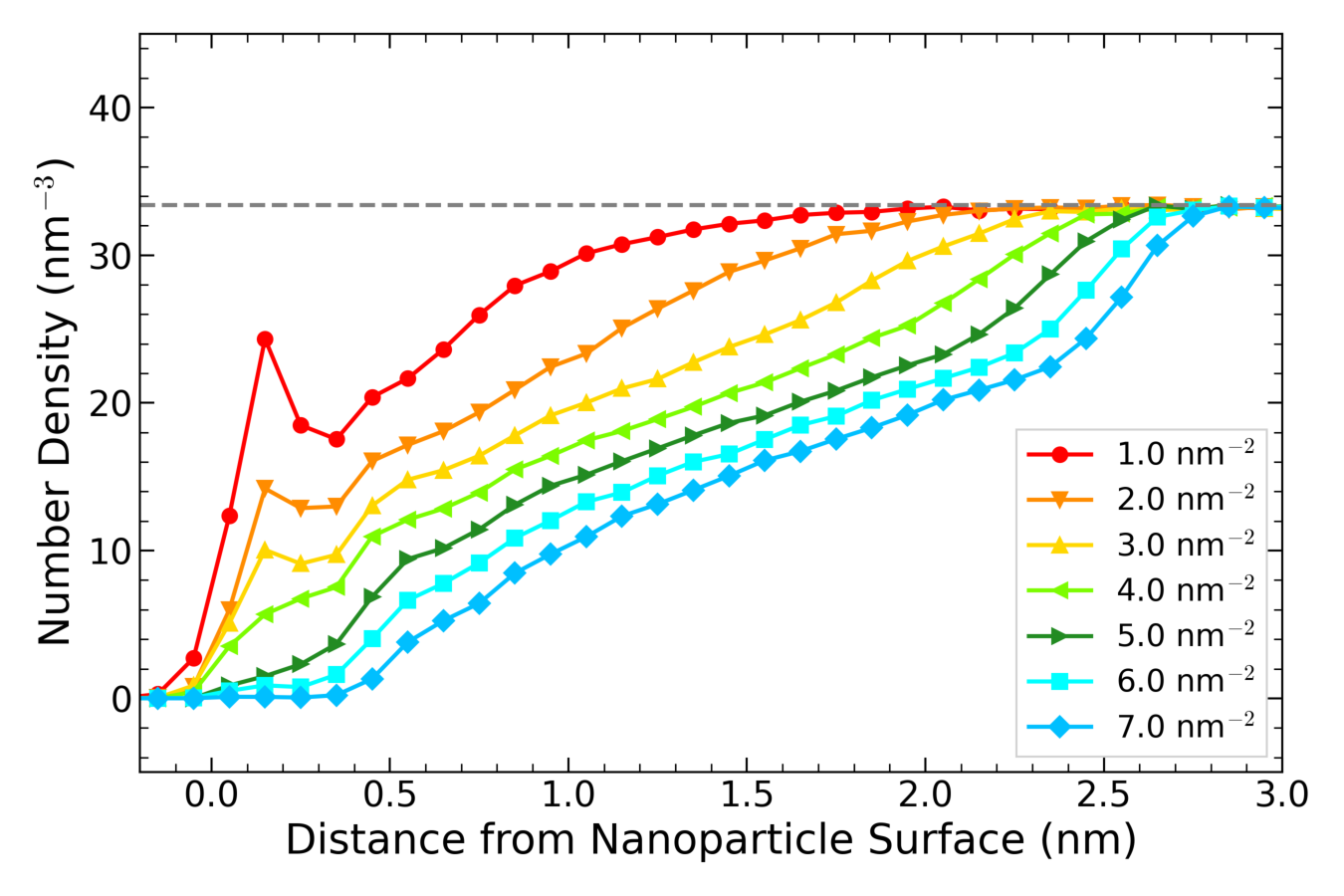}
    \caption{Number density of water molecules in the vicinity of the NP coated with thiol-PEG$_8$-amine ligands in surface densities ranging from 1.0 nm$^{-2}$ to 7.0 nm$^{-2}$. The grey dashed line represents the number density of water at ambient conditions, 33.4 nm$^{-3}$.}
    \label{fig:3.7nmWaterPlot}
\end{figure}

Analysis of water distribution near the surface of metal NPs is one of the important aspects for evaluating the ability of coated metal NPs to act as radiosensitising agents. Figure~\ref{fig:3.7nmWaterPlot} presents the number density of water molecules near the surface of a 3.7~nm NP coated with thiol-PEG$_8$-amine ligands at ligand surface densities $\sigma = 1.0 - 7.0$~nm$^{-2}$. In the region close to the NP surface, the number density distribution of water molecules differs significantly for the different values of ligand surface density. At the lowest ligand density considered, the number density distribution of water molecules has a maximum, and the amplitude of this peak decreases with increasing ligand surface density. Such behaviour has a simple geometrical explanation: As ligand density increases, there is no space left for water molecules to penetrate the coating region toward the NP surface. The amount of water in proximity to the NP surface is essential for the production of hydroxyl radicals \cite{HydroxylRadicalProduction, Haume2018}, which may have a strong impact on the radiosensitising effectiveness of metal NPs.

\begin{figure}[t!]
    \centering
    \includegraphics[width=0.48\textwidth]{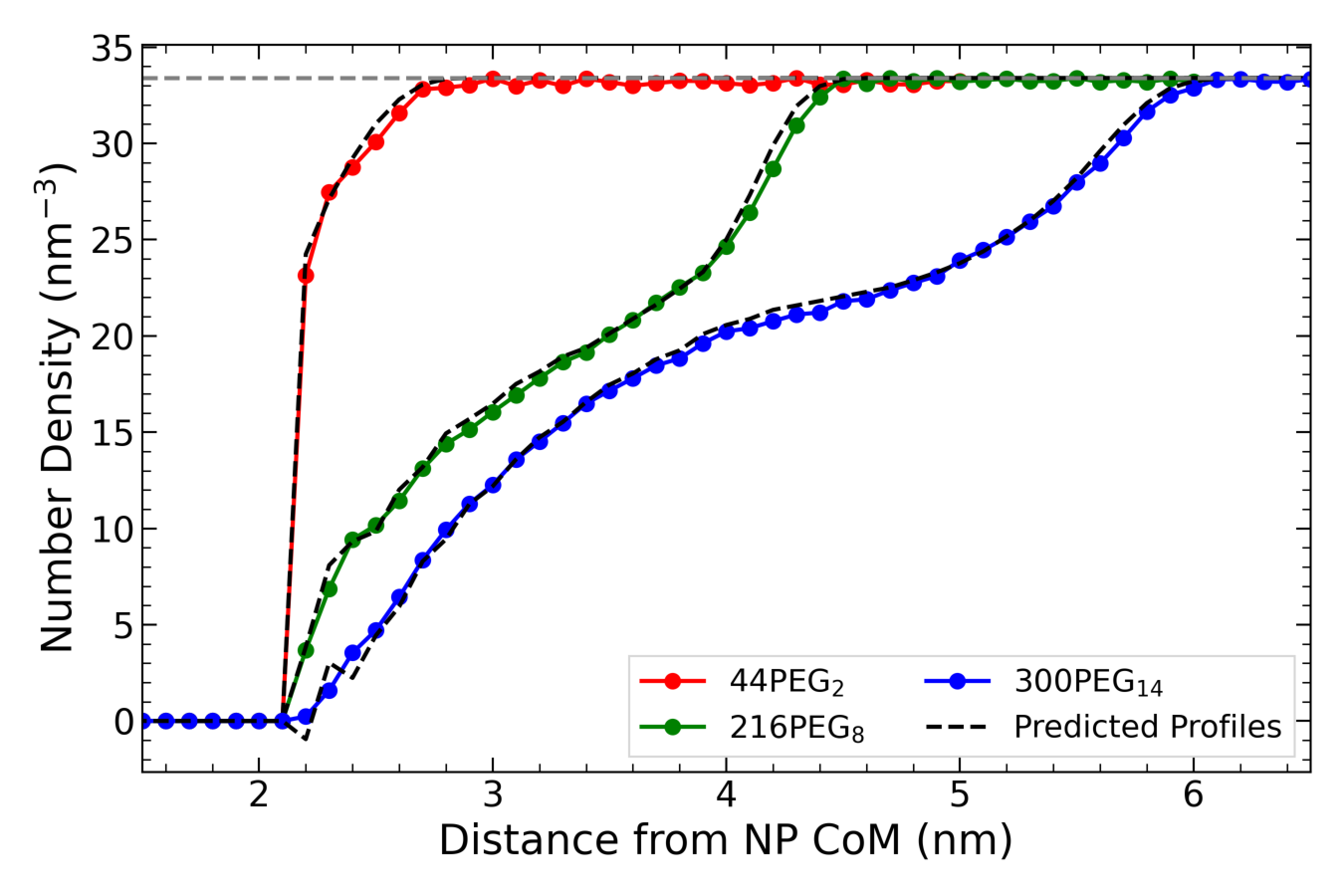}
    \caption{Radial distributions of the number density of water molecules in the vicinity of 3.7~nm NPs coated with 44 thiol-PEG$_2$-amine, 216 thiol-PEG$_8$-amine, and 300 thiol-PEG$_{14}$-amine ligands at surface densities of 1.0~nm$^{-2}$, 5.0~nm$^{-2}$, and 7.0~nm$^{-2}$ respectively. Results from the MD simulations performed in this study are shown as coloured symbols; the number densities of water estimated by means of Eq.~(\ref{eqution:WaterConversion}) are shown by dashed black curves. Dashed grey line is the number density of water at ambient conditions.}
    \label{fig:Water_PEG_Comparison_Plot}
\end{figure}

For the systems considered in this study, the coating region around the metal NP comprises only thiol-PEG-amine and water molecules. Therefore, by determining the radial dependence of the number density of thiol-PEG-amine, one can obtain the number density of water based on the excluded volume principle \cite{Hermans_1982_JCP.77.2193}:

\begin{equation}
    n_{\text{water}}(r) = \left( 1 - \frac{n_{\text{PEG}}(r)}{n_{\text{PEG}}^{\text{max}}} \right) n_{\text{water}}^{\text{a}} \ .
    \label{eqution:WaterConversion}
\end{equation}
Here $n_{\text{PEG}}(r)$ is the atomic number density of thiol-PEG-amine ligands as a function of the radial distance from the CoM, $n_{\text{PEG}}^{\text{max}}$ is the maximum number density of thiol-PEG-amine given by Eq.~(\ref{equation:MaxNumberDensity}), and $n_{\text{water}}^{\text{a}} = 33.4$~nm$^{-3}$ is the number density of water molecules at ambient conditions. Figure~\ref{fig:Water_PEG_Comparison_Plot} compares the number density profile of water molecules determined from MD simulations (symbols) with that evaluated using Eq.~(\ref{eqution:WaterConversion}) (dashed curves).
The evaluated water density profile depends on the maximum number density of ligands, $n_{\text{PEG}}^{\text{max}}$, estimated using Eq.~(\ref{equation:MaxNumberDensity}). The latter, in turn, depends on $r_0$, the cross-sectional radius of a cylinder surrounding a ligand molecule, which should approximately be equal to a characteristic van der Waals radius of atoms of the ligands. Setting this value to $r_0 = 0.26$~nm provides a good correspondence to the simulation results. Therefore, the main features of the water number density profile obtained from the atomistic MD simulations can be evaluated with good accuracy by means of Eq.~(\ref{eqution:WaterConversion}).

\section{Conclusion}
\label{sec:Conclusions}

This study has been devoted to atomistic modelling and structural characterisation of a 3.7~nm gold NP coated with thiol-PEG$_n$-amine molecules of different lengths $(n = 2-14)$ and solvated in water. The system size and composition have been selected in connection to several radiosensitisation experiments \cite{Sophie, GU2009196}.
The coating structure and water distribution near the NP surface have been characterised on the atomistic level using the MBN Explorer \cite{MBNExplorer} and MBN Studio \cite{MBNStudio} software packages. It has been found that the radial distribution of the number density for coating molecules of different lengths and surface densities is general between different NP sizes and can be divided into three spatial regions. The first region is characterised by the distribution of ligands over the NP metal surface and, therefore, the surface geometry and size of the NP. The second region is characterised by the distribution of the carbon and oxygen atoms of the PEG backbone and is characterised by inverse square dependence on the distance from the NP center of mass. The third region is characterised by the specific conformation state of the thiol-PEG-amine coating, with ligands acquiring either ``mushroom''-like or ``brush''-like shapes depending on the surface density of ligands attached to the NP.

The results of MD simulations presented in this study, in combination with the results of our recent study \cite{Verkhovtsev_2022} and those from the field of polymer physics, have been used to calculate key structural parameters of the coatings of radiosensitising gold NPs and establish connections between the coating structure and distribution of water for different NP sizes as well as lengths and surface densities of coating molecules.
The results of the quantitative analysis of water distribution in the vicinity of coated metal NPs can be used in follow-up studies to evaluate the radiosensitising effectiveness of a particular NP system based on the proximity of water to the NP metal core, which should impact the production hydroxyl radicals and reactive oxygen species in the vicinity of metal NPs.

\begin{acknowledgments}
The authors acknowledge financial support from the European Union's Horizon 2020 research and innovation programme – the RADON project (GA 872494) within the H2020-MSCA-RISE-2019 call and from the COST Action CA20129 MultIChem, supported by COST (European Cooperation in Science and Technology).
The possibility of performing computer simulations at the Goethe-HLR cluster of the Frankfurt Center for Scientific Computing is gratefully acknowledged.
\end{acknowledgments}

\bibliography{sn-bibliography.bib}

\end{document}